\documentclass[runningheads]{llncs}
\usepackage{graphicx}
\usepackage{graphicx}
\usepackage{caption}
\usepackage{hyperref}
\usepackage{amsfonts}
\usepackage{amssymb}
\usepackage{amsmath}
\usepackage{xcolor}
\usepackage[numbers]{natbib}
\usepackage{float}
\usepackage{adjustbox}
\usepackage{tabularx}
\usepackage{epstopdf}
\usepackage{booktabs}
\usepackage{comment}
\usepackage{multirow}
\usepackage{breqn}
\usepackage{inputenc}
\usepackage{dblfloatfix}
\usepackage{appendix}
\usepackage{footnote}
\usepackage{threeparttable}
\makesavenoteenv{tabular}
\begin{document}
\title{A Policy-oriented Agent-based Model of Recruitment into Organized Crime  \thanks{This contribution originates within the framework of PROTON project. The PROTON project has received funding from the European Union’s Horizon 2020 research and innovation programme under grant agreement Nr. 699824. We want to thank Nicolas Payette, who contributed to the development of a previous version of the model.}}
%
%
\author{Gian Maria Campedelli\inst{1}\orcidID{0000-0002-7734-7956} \and
Francesco Calderoni\inst{1}\orcidID{0000-0003-2979-4599} \and
Mario Paolucci\inst{2}\orcidID{0000-0002-8276-1086} \and
Tommaso Comunale\inst{1}\orcidID{0000-0001-6820-784X} \and
Daniele Vilone\inst{2}\orcidID{0000-0002-3485-9249} \and
Federico Cecconi\inst{2} \and
Giulia Andrighetto\inst{2,3,4}\orcidID{0000-0002-3896-1363}} 
\authorrunning{G.M. Campedelli et al.}
%
\institute{ Transcrime - Universita Cattolica del Sacro Cuore, 20123 Milan, Italy \email{gianmaria.campedelli@unicatt.it}\and
National Research Council - Institute of Cognitive Sciences and Technologies - Laboratory of Agent-based Social Simulation, 00185 Rome, Italy \and
Institute for Future Studies, SE-101 31 Stockholm, Sweden \and
M\"alardalen University, 722 20 V\"asteras, Sweden}
\maketitle              
\begin{abstract}
Criminal organizations exploit their presence on territories and local communities to recruit new workforce in order to carry out their criminal activities and business. The ability to attract individuals is crucial for maintaining power and control over the territories in which these groups are settled. This study proposes the formalization, development and analysis of an agent-based model (ABM) that simulates a neighborhood of Palermo (Sicily) with the aim to understand the pathways that lead individuals to recruitment into organized crime groups (OCGs). Using empirical data on social, economic and criminal conditions of the area under analysis, we use a multi-layer network approach to simulate this scenario. As the final goal, we test different policies to counter recruitment into OCGs. These scenarios are based on two different dimensions of prevention and intervention: (i) primary and secondary socialization and (ii) law enforcement targeting strategies. 

\keywords{Criminal Organizations  \and Disruption Strategies \and Public Policy \and Criminal Involvement \and Law Enforcement }
\end{abstract}
\section*{Introduction}
Organized crime (OC hereinafter) is present in a wide number of countries all over the world \cite{FijnautOrganisedCrimeEurope2004, VareseMafiasMoveHow2011, AthanassaopolouOrganizedCrimeSoutheast2013, AbadinskyHistoryOrganizedCrime2014, WilliamsNigerianCriminalOrganizations2014}. Besides the peculiar differences between different criminal organisations, they all pose social, economic and security challenges to societies, institutions and legal economies. OCGs are capable to attract people and strengthen their presence on a territory. Human workforce is a crucial asset for OCGs: a higher number of affiliates and members is an indicator of the actual resources at their disposal and a measure of the potential strength of the groups itself. Understanding how recruitment works, then, becomes a relevant issue both from the research and the policy standpoints. Scientific inquiry can help in pursuing this goal. However, investigating recruitment dynamics at scale is extremely costly in a real-world setting, involving first and foremost feasibility constraints. Additionally, it also poses ethical questions that are often present when dealing with social experiments. Computer simulations, conversely, can overcome these issues. With this regard, their potential has also started to be investigated in criminology \cite{Brantinghamcomputationalmodelsimulating2005, MallesonCrimereductionsimulation2010, TroitzschExtortionRacketsEventOriented2016, NardinSimulatingprotectionrackets2016}. 
\\
In light of these considerations, this paper presents the rationale and structure of an agent-based model of recruitment into OC. To best of our knowledge, this is the first simulation model that specifically integrates computational science and criminology to address the problem of recruitment into OC. The ABM will resemble a neighbourhood of Palermo, the main city of Sicily: our artificial society will comprise 10,000 agents with socio-economic, demographic and criminal characteristics derived from empirical data. The society will be modelled through a multiplex network structure. The investigation of recruitment dynamics will be coupled with the final goal of the simulation, which is the testing of potential policies to prevent or reduce recruitment into OCGs, especially for youth. 
\\
The rest of the paper is organised as follows: the theoretical framework part will briefly cover the main  theories that constitute the criminological backbone of the model. The Data section will describe the empirical information used in the ABM and the sources from which data have been extracted. The Main Structure section will thoroughly explain the rationale of the multiplex approach and the two main components of the model. Finally, in the Policy Scenarios section, the two proposed families of counter-policies that respectively deal with socialisation and network disruption will be presented.

\section{Theoretical Framework}
The rationale of our simulated society is rooted in the outcomes of a recent systematic review on the factors leading to recruitment into OC \cite{SavonaSystematicreviewsocial2017}. Furthermore, it also takes into account several theoretical perspectives. Some of these suggest that organised crime is embedded in the social environment and that social relations are crucial for the recruitment into organised crime. With this regard, differential association theory and social learning theory \cite{SutherlandProfessionalThief1937, BurgessDifferentialAssociationReinforcementTheory1966, BruinsmaDifferentialAssociationTheory2014} posit that crime in its various aspects is learned in a social environment by relating with other criminal agents. Empirical studies have also highlighted that OC is socially and criminally embedded in the surrounding environment \cite{GranovetterEconomicActionSocial1985,McCarthyGettingStreetCrime1995}. The position agents occupy within a criminal network determines their possibilities to commit crimes. In this sense, an agent's valuable criminal ties determine his social opportunity structure \cite{KleemansCriminalCareersOrganized2008}.
Conversely, other theories such as the general theory of crime \cite{Gottfredsongeneraltheorycrime1990} argue that an individual's low self-control levels determine an inability to compute the negative consequences of one's criminal behaviour, thereby determining persisting patterns of criminality throughout his life. The general theory of crime contends that group crime does not have specific characteristics and that the formation of criminal groups is mostly driven by self-selection processes.
The social relations (i.e., social learning and differential association) and self-control (i.e., general theory of crime) perspectives may generate opposing views about the recruitment into organised crime: however, a more comprehensive explanation of criminal activity could be reached via the combination of elements of both frameworks. With this regard, the development of an agent-based model is a convenient way to do so, since its flexibility can allow to integrate both personal and inter-personal components. In light of this, the present model operationalizes criminal involvement both as the result of interaction with others and as emerging from agents’ individual characteristics pushing towards crime.

\section{The Model}
\subsection{Data}

Using empirical data to feed the simulations is fundamental when aiming at setting up a reasonable and grounded model, besides theoretical and formal mechanisms (e.g., the mechanism of crime commission). Furthermore, data shall also be used ex-post to assess whether the model produces reliable and plausible results, especially considering the policy-oriented objectives of the ABM.
To develop the model and validate the results, we have retrieved and processed several data from different sources regarding specific demographic, economic, social and criminal aspects. We have chosen the city of Palermo as the specific setting to be resembled by the simulation model, as Palermo is one of the cities with the highest mafia presence in Italy \cite{DugatoMeasuringOrganisedCrime2019a} (Table \ref{data}).

\subsection{A Multiplex-Network Approach}
Simulating the dynamics and processes that lead to the recruitment into OC requires to take into account a wide variety of factors. While certain elements are inherently linked to the individual sphere (e.g. age, gender), others span over the personal characteristics of an agent: making new friends, for instance, is dependent upon the social environment in which an agent is set. Two children at the elementary school are more likely to become friends if they are in the same classroom or if they have the same age, rather than being separate into different classrooms or belong to different years.
\newpage

\begin{table}[]
\caption{Data Employed to Inform and Validate the Simulation}
\centering
\footnotesize
\begin{tabular}{lll}
\hline\hline
\textbf{Data} & \textbf{\begin{tabular}[c]{@{}l@{}}Variable\\ Type\end{tabular}} & \textbf{\begin{tabular}[c]{@{}l@{}}Source,\\ Time Span\end{tabular}} \\ \hline
\begin{tabular}[c]{@{}l@{}}Distribution of female fertility\\ according to age\end{tabular} & Demographic & Istat, 2017 \\ \hline
\begin{tabular}[c]{@{}l@{}}Mortality probability \\ by age and gender\end{tabular} & Demographic & Istat, 2016 \\ \hline
\begin{tabular}[c]{@{}l@{}}Wealth level distribution by \\ a person's level of education\end{tabular} & Economic & \begin{tabular}[c]{@{}l@{}}Bank of Italy Survey, 2016\\ Istat, 2011\end{tabular} \\ \hline
\begin{tabular}[c]{@{}l@{}}Distribution of employer sizes\\ in Palermo\end{tabular} & Economic & Istat, 2011 \\ \hline
\begin{tabular}[c]{@{}l@{}}Household size distribution\\ by the household head's age \\ in Palermo\end{tabular} & Demographic & \begin{tabular}[c]{@{}l@{}}Census data, Istat 2011, \\ Municipality of Palermo\end{tabular} \\ \hline
\begin{tabular}[c]{@{}l@{}}Distribution of household\\ sizes in Palermo\end{tabular} & Demographic & \begin{tabular}[c]{@{}l@{}}Census data, Istat 2011, \\ Municipality of Palermo\end{tabular} \\ \hline
\begin{tabular}[c]{@{}l@{}}Distributionof household type \\ by age of household head\end{tabular} & Demographic & \begin{tabular}[c]{@{}l@{}}Census data, Istat 2011, \\ Municipality of Palermo\end{tabular} \\ \hline
\begin{tabular}[c]{@{}l@{}}Distribution of people's age\\ by gender in Palermo\end{tabular} & Demographic & Istat, 2018 \\ \hline
\begin{tabular}[c]{@{}l@{}}Number of schools in the city\\ of Palermo by level of\\ education\end{tabular} & Social & \begin{tabular}[c]{@{}l@{}}Ministry of Education\\ and Research, 2016\end{tabular} \\ \hline
\begin{tabular}[c]{@{}l@{}}Distribution of Mafia families \\ in Palermo\end{tabular} & Criminal & \begin{tabular}[c]{@{}l@{}}Corte d'Appello di \\ Reggio Calabria, 2012;\\ Criminal Investigations \\``Aemilia",  ``Crimine'', \\ ``Infinito'', ``Minotauro''\end{tabular} \\ \hline
Co-offending prevalence & Criminal & Istat, 2012-2016 \\ \hline
\begin{tabular}[c]{@{}l@{}}Crime rates\\ (corrected for dark number)\end{tabular} & Criminal & Istat, 2012-2016 \\ \hline
Punishment distribution & Criminal & Istat, 2012-2016 \\ \hline
Imprisonment length distribution & Criminal & Istat, 2012-2016 \\ \hline
\end{tabular}
\label{data}
\end{table}

\begin{figure}[!th]
    \centering
    \includegraphics[width=0.3\textwidth]{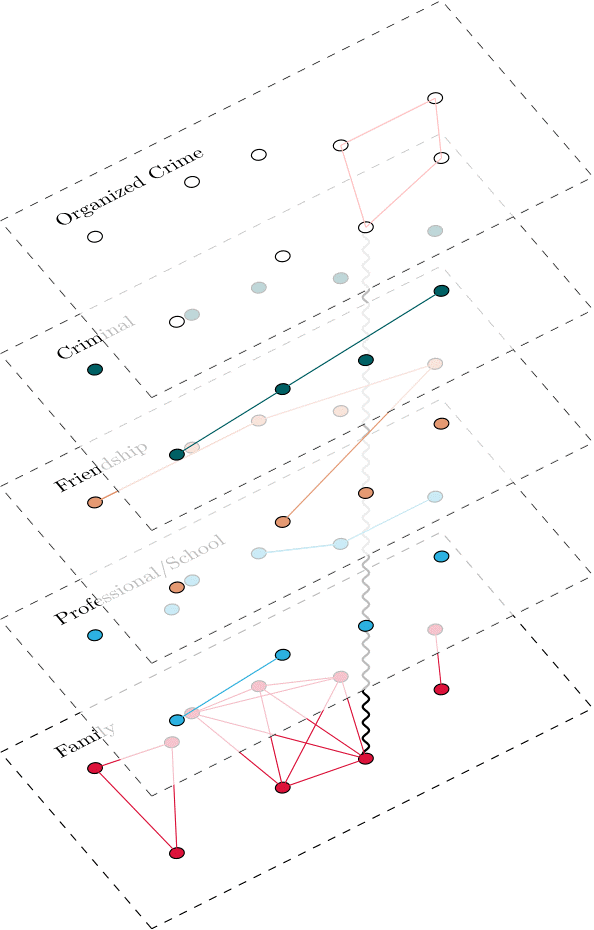}
    \caption{Graphic Depiction of the Multiplex Network Structure}
    \label{nets}
\end{figure}

In real life, every person engages in different types of relations, e.g. as part of a family, in friendships, at work, and - if criminals - in co-offending. A member of an OCG is also part of a wider social environment as embedded in multiple social worlds. The literature proves that relations of different types may drive the involvement and recruitment into organised crime \cite{Arlacchimafiaimprenditrice1983, ArsovskaDecodingAlbanianorganized2015, Brancaccioclandicamorra2017}.
To adequately address the dynamics of individual and social drivers, we opted for an ABM based on a multiplex rationale. A multiplex network includes several networks, each mapping specific social relations.  Five relational layers are modelled in the simulations: families, friendship networks, professional and school ties, criminal relations and organised crime groups (Figure \ref{nets}). 

Considering the need for creating a society that synthetically mirrors the real-world, we have decided to include all the main dimensions through which an individual can realistically act and behave. The structure, topology and characteristics of the networks are empirically grounded using official statistics or replicating mechanics found in existing scientific works. The multiplex network framework also allows for considering individual-level characteristics as agents’ attributes, thus making possible to analyse individual and social factors to simulate realistic recruitment dynamics. Agents in the simulation, regardless of being or not part of an OC groups, can be born, get engaged/married, make children, die, create and break relations, and commit crimes. \\
Specifically focusing on the organised crime dimension, the model considers one single OCG existing in the simulation. Its internal structure, composition (in terms of gender distribution and generation/age distribution) reflects the ones found through analyses of several police investigations on Italian OC groups. 

\subsection{Recruitment into Organised Crime}
\begin{figure}[!th]
    \centering
    \includegraphics[width=0.6\textwidth]{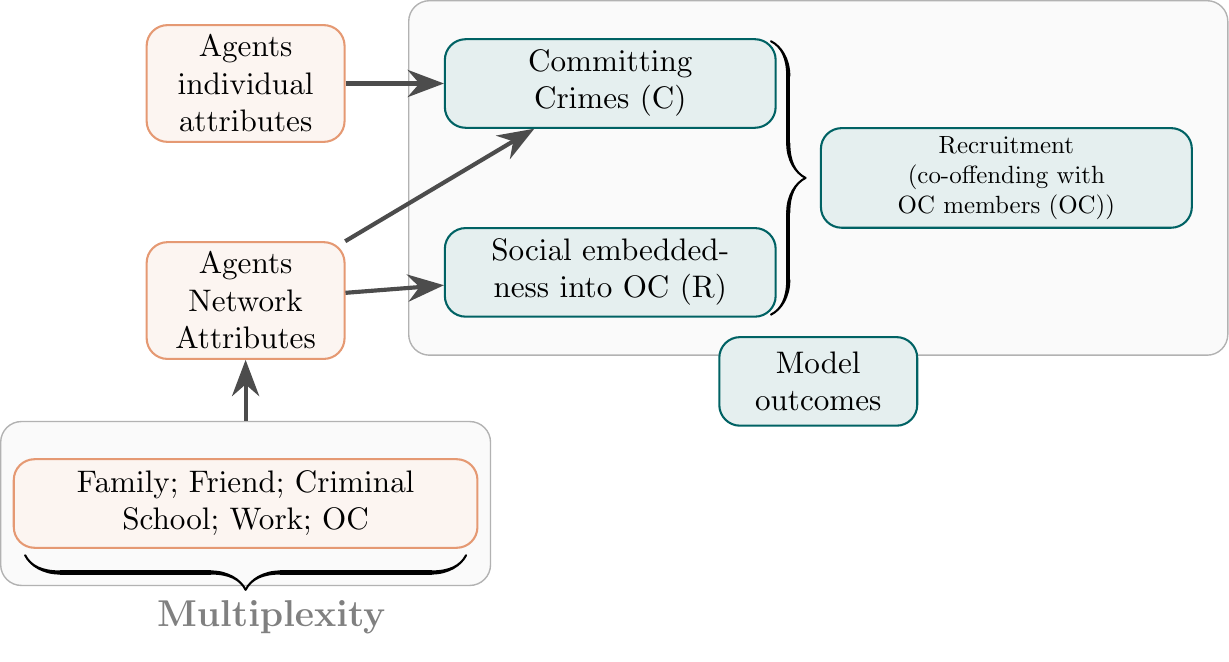}
    \caption{Synthetic Description of the General Structure of the Model}
    \label{net}
\end{figure}

For the purpose of this simulation, recruitment occurs when an agent commits a crime with at least another agent who is already a member of the OCG.\footnote{It has to be noted that the model is initialised with a certain number of agents estimated based on data retrieved from criminal investigation, as showed in Table \ref{data}.} This option was driven by different considerations. First, it is observable and easily operationalizable. Requiring the commission of a crime with OCG members models in a straightforward way the process of recruitment, avoiding subjective evaluations. Second, it is broadly consistent with the criminal law approaches criminalising organised crime across countries \cite{CalderoniOrganizedCrimeLegislation2010}.
Two different complementary dimensions contribute to the determination of the recruitment processes in the model, namely the probability of committing a crime (called $C$) and the embeddedness into organised crime (called $R$) (Figure \ref{net}).

\subsubsection{Modelling Crime Commission: the ``$C$'' Function}
The $C$ function models the probability that agent $i$ will commit a crime at time $t$. Its structure revolves around two components: the probability of committing a crime given gender and age class and the risk of committing a crime associated with several social and criminal factors. The first component has been derived estimating baseline probabilities of committing a crime in Sicily using official statistics. The probabilities are computed using both the figures on reported crimes and data gathered from victimisation surveys. This complementary computation allows to eliminate the problem of dark figure. The dark figure refers to the difference between the real number of crimes and the actual reported number of crimes \cite{SkoganDimensionsDarkFigure1977a}. In the model, we have thus reliably estimated the actual real number of crimes occurred in Palermo in the period 2012-2016. Baseline probabilities are reported in Table \ref{baseline}.

\begin{table}[]
\centering
\caption{Gender and age class probabilities ($(C|\theta(g,a)$) and Odds Ratios of committing a crime in Palermo}
\footnotesize
\begin{tabular}{lcc}
\hline
\textbf{(Sex, Age Class)} & \textbf{Probability} & \textbf{OR} \\ \hline\hline
(Female,  $\leq${}13 yrs) & 0.0004 & 0.0004 \\ \hline
(Female, 14-17) & 0.0223 & 0.0229 \\ \hline
(Female, 18-24) & 0.0511 & 0.0538 \\ \hline
(Female, 25-34) & 0.0634 & 0.0677 \\ \hline
(Female, 35-44) & 0.0643 & 0.0687 \\ \hline
(Female, 45-54) & 0.0489 & 0.0514 \\ \hline
(Female, 55-64) & 0.0308 & 0.0318 \\ \hline
(Female, $\geq$ 65) & 0.0111 & 0.0112 \\ \hline
(Male, $\leq${}13 yrs) & 0.0022 & 0.0022 \\ \hline
(Male, 14-17) & 0.1502 & 0.1767 \\ \hline
(Male, 18-24) & 0.3019 & 0.4324 \\ \hline
(Male, 25-34) & 0.3036 & 0.4359 \\ \hline
(Male, 35-44) & 0.2751 & 0.3795 \\ \hline
(Male, 45-54) & 0.1996 & 0.2494 \\ \hline
(Male, 55-64) & 0.1268 & 0.1453 \\ \hline
(Male, $\geq$ 65) & 0.0537 & 0.0567 \\ \hline
\end{tabular}
\label{baseline}
\end{table}
The second component originates from empirical findings in the existing literature. Specifically, several systematic reviews providing information on the impact of certain factors on the general probability of committing a crime. These sources provide effect sizes (in different forms, e.g. odds ratios) allowing to determine the different probabilities of coming a crime given an agent’s network and individual characteristics. The list of individual factor-based rules that drive the model rules for committing a crime is presented in Table \ref{factor}.

\begin{table}[]
\caption{Individual-level Factors ($\gamma$)  Driving the Crime Commission Process - Odds Ratios}
\footnotesize
\setlength{\tabcolsep}{10pt}
\centering
\begin{threeparttable}
\begin{tabular}{p{2.8cm} c p{5cm}}
\hline
\textbf{Risk Factor} & \textbf{OR} & \textbf{Definition} \\ \hline\hline
Unemployment & 1.30 & Having/not having a job  \\ \hline
Education & 0.94 & Having/not having an high school diploma \\ \hline
Natural Propensity   & 1.97 & Having a crime propensity higher than a certain value $x$ (log-normally distributed in the population) \\ \hline
Criminal History  & 1.62 & Having/not having committed a crime in the past \\ \hline
Criminal Family  & 1.45 & Having a share of criminal family ties which is higher or equal to 0.5.\tnote{*} \\ \hline
Criminal Friends and Co-Workers & 1.81 & Having a share of criminal friends ties which is higher or equal to 0.5.\tnote{**}\\ \hline
OC Membership & 4.5 & Being part of an OC group \\ \hline
\end{tabular}
\begin{tablenotes}\footnotesize
\item[*] \scriptsize A criminal family tie is a direct link with a family member which has committed at least one crime in the last 2 years.
\item[**] A criminal friendship/professional tie is a direct link with a family member which has committed at least one crime in the last 2 years.
\end{tablenotes}
\end{threeparttable}
\label{factor}
\end{table}

Given the data at our disposal, we model the probability of committing a crime $p(\bar{C})$ for an individual $i$ at time $t$ as: 

\begin{equation}
    p(\bar{C})_{i,t}=\left [ (C|\theta(g,a)_{i,t})\left ( \sum_{j=1}^{m} \gamma_{j_{i,t}} \right )\right ]+\varepsilon
\end{equation}
where $(C|\theta(g,a)_{i,t})$ is the baseline probability for the individual given its gender and age class, and  $(\sum_{j=1}^{m} \gamma_{j_{i,t}})$ is the summation of the risk factors $\gamma$ and $\varepsilon$ is an error term stochastically distributed in order to bound the individual probabilities of committing a crime to the population average.  
Specifically, given the odds ratio of a risk factor, we increase/decrease the baseline risk by the percentage provided by the Odds Ratio (OR) itself (e.g.: if the OR is equal to 1.41, and an individual has it among their characteristics, and their baseline is 0.15, it means that the final value has to be the product between the baseline and 0.41, namely the increase of the risk in percentage given that risk factor). Therefore, at each time of reference $\bar{t}$ and for each subset of the population $(g,a)$ of given gender and age class, the following equation shall hold:
\begin{equation}
C(g,a)_{\bar{t}}\approx \frac{1}{n(g,a)}\sum_{i=1}^{n(g,a)}p(\bar{C})
\end{equation}
The equation means that at each time of reference, the average probability of committing a crime for all individuals belonging to the same (gender,age) class shall be approximately similar to the fixed average values presented in Table 1, where approximately means that we can allow the model to float in a $\pm$ 0.1 range in order not to set overly deterministic mechanics to the model. In other words, we model the distribution of $C$ as a strictly stationary and ergodic random process:

\begin{equation}
    F_{C}\left [C(g,a)_{t_{1}}, ... \; , C(g,a)_{t_{k}} \right ]=F_{C}\left [C(g,a)_{t_{1+\tau }} , ... \;, C(g,a)_{t_{k+\tau}} \right ] \mathrm{for} \; \forall \; \tau, t_{1},...,t_{k}
\end{equation}
$C$ has been computed to provide realistic figures on committed offences within the model, in the form of rates by 100,000 inhabitants, using official statistics for different years (2012-2016, specifically). The calculation relied on the correction of crime figures by the dark number of each crime category. This allowed to take into account and include also those offences that have not been discovered or prosecuted in the original data at our disposal, thus giving more solid and reliable estimates.
\\
Most crimes are committed by single offenders \cite{StolzenbergCoOffendingAgeCrimeCurve2008, vanMastrigtCoOffendingAgeGender2009, CarringtonCooffendingCanadaEngland2013}. Based on the literature on co-offending, a few crimes will require more than one offender. These works also help in shaping the mechanisms that lead two or more individuals to commit a crime together. Peer and more in general social influence play a relevant role in driving this criminal cooperation  \cite{WeermanTheoriesCooffending2014}. The model thus matches co-offenders based on mechanisms of social proximity: the closer two agents are in terms of social relations across network layers and the higher is the value of $C$ of both individuals, the higher the probability of becoming co-offenders.\\
Once agents commit crimes they can be incarcerated. Incarceration is estimated using empirical data retrieved from official statistics. Apart from family links, an agent in prison loses all the ties that he/she has created during his/her life (including during his/her job). The mechanism for incarceration is based on a countdown that allows to establish when the agent leaves prison and returns to be “free” in the society, recovering part of its ties.

\subsubsection{Modelling Organized Crime Embeddedness: Defining ``$R$''.}
$R$ defines the embeddedness into OC and allows to adequately model the probability of an artificial agent to be recruited into a criminal organisation. $R$ seeks to consider the role of each agent’s social community in the process of recruitment. The theoretically-driven assumption is that individuals that are embedded in communities (across all types of networks considered by our simulation) that are highly populated by OC members face higher risks of being recruited. R affects the selection of new OC members in the simulation. For example, among two equally suitable co-offenders, OC members are likely to co-offend with the agent who is more embedded in OC. In a simple form - but coherently with the differential association and social learning theories as well as the social embeddedness of OC - $R$ is then operationalised as the proportion of OC members among the social relations of each individual (comprising family, friendship, school, working and co-offending relations). 
\\
In mathematical notation, a multiplex network $\mathcal{G}=\left \{G^{1},...,G^{l},...,G^{M} \right \}$ which resembles our simulated society is a set of $M$ single-layer networks that get dynamically updated at each time unit $t$. Each single-layer network is denoted as $G^{l}=(V,E)$, that takes the form of a $N \times N$ matrix. Given this notation, for each $G^{l}$, we define an $h$-hop neighbourhood graph for each node $i$. The node set of the $h$-hop neighbourhood graph is defined as the set $N^{h}_{i}=\left \{ j|k \in N^{h-1}_{i}, j\in V, (k,j) \in E \right \}\cup N^{h-1}_{i}$ with $h\geq1$. The set of edges is then formalised as $E^{h}_{i}=\left \{ (j,k)|j \in N^{h-1}_{i}, k \in N^{h}_{i} , (j,k) \in E) \right \}$. The local neighbourhood of agent $i$ in the single layer network $G^{l}=(V,E)$ becomes then a vector  $\mathbf{w}^{G^{l}}_{i}=\left[ \mathrm{w}_{i,l}^{G^{l}}\cdots \mathrm{w}_{i,j}^{G^{l}}\right ]$ where each element represents the weight of the edges included in the $h$-hop local neighbourhood of the agent. Each value of the vector follows the relation $\mathrm{w} \propto h^{-1}$, meaning that the weights are inversely proportional to the distance between the ego $i$ and an agent $j$ included in the $h$-hop network.
At this point, to compute the embeddedness $R$ of an agent $i$ in his local community, we sum over the vectors of each single-layer network:

\begin{equation}
\mathbf{w}^{\mathcal{G}}=\left [ \mathrm{w}_{i,l}^{G^{1}}\cdots \mathrm{w}_{i,j}^{G^{1}}\right ]+\cdots+ \left[ \mathrm{w}_{i,l}^{G^{l}}\cdots \mathrm{w}_{i,j}^{G^{l}}\right ]=\left [ \mathrm{w}^{\mathcal{G}}_{i,l} \cdots \mathrm{w}^{\mathcal{G}}_{i,j} \right ]
\end{equation}
This equation yields the resultant vector of weights deriving from the complete agent's $h$-hop network. To calculate the actual OC embeddedness, we derive the resultant vector of weights obtained from the agent's $h$-hop OC network $\Theta^{\mathcal{G}}_{i}=\left [ \theta^{\mathcal{G}}_{i,l} \cdots \theta^{\mathcal{G}}_{i,j} \right ] $, such that the node set is called $N^{h}_{i\: OC}$ and the set of edges is $E^{h}_{i\: OC}$, where $N^{h}_{i\: OC}\subseteq N^{h}_{i}$ and $E^{h}_{i\: OC}\subseteq E^{h}_{i}$.  $R$ is finally mathematically defined as:

\begin{equation}
R_{i}=\frac{\sum_{i=1}^{N^{h}_{OC}} \theta^{\mathcal{G}}_{i,j}}{\sum_{i=1}^{N^{h}} \mathrm{w}^{\mathcal{G}}_{i,j}} \: \in [0,1]
\end{equation}
which is the ratio between the total number of weights in the OC $h$-hop network and the general $h$-hop network of agent $i$. The values of $R$ fall in the range [0,1], with 1 indicating complete overlapping between the general $h$-hop networks and 0 highlighting total absence of OC members in the local community of the agent.  The proposed method implicitly weights the OC embeddedness such that (i.) the importance of OC ties is inversely proportional to the distance and (ii.) the importance of OC ties (but also of other non-OC ties) is proportional to the number of different ties between any two individuals.

In addition to the contribution in determining recruitment, $R$ provides useful information also to analyse the general dynamics of the model. First, it enables to clearly distinguish between active OC members and pro-OC agents. An agent may be strongly embedded in OC-prone networks but not necessarily be a member. For example, this could be the case of women who are certainly living in OC-prone contexts (e.g. wives and daughters of OC members)  but are rarely charged and convicted as OC members since they generally refrain from committing offences. Similarly, a juvenile son of an OC member who is just two years old cannot be considered an active member, but would still have a very high value of $R$, making it very likely that he will be recruited in the future. Second, R may contribute to the simulation of prevention policies, especially those on the primary and secondary socialisation. The simulation will need to identify the target population and R could contribute in identifying the population at risk better than merely relying on other indicators e.g. the number of crimes committed by the parents or the involvement of a parent into OC.

\section{Policy Scenarios}
The  policy scenarios that have been selected constitute the final goal of the ABM and they have followed suggestions and remarks given by policymakers belonging to several European institutions (e.g., Europol, Dutch Ministry of Justice, Italian Ministry of Interior). They are specifically divided into two distinct types: (i) primary and secondary socialisation and (ii) law enforcement. 
\subsection{Primary and Secondary Socialisation}
This policy scenario aims at protecting juveniles from socialisation processes leading them to recruitment into OC. Among the possible solutions, policies aiming at decreasing or countering the influence of OC-prone social relations may offer an effective strategy to prevent the recruitment into OC. These policies will be modelled through different solutions. They include (a) reducing the influence of parents (normally the father), (b) addition of pro-social ties such as non-criminal friends, (c) provision of educational and job opportunities (increasing education achievement, wealth and work ties). This scenario will put to the test and compare the effectiveness of these strategies both as separate and combined treatments.\\
The primary socialisation policy targets young people aged 12-18 living in OC families, intended as OC families, families where at least one parent is an OC member. It will be possible to select the share of the target young people based on a risk score reflecting the family embeddedness ($R$) in OC and/or based on the OC members convicted in the simulation. This group of juveniles at risk will then be subjected to interventions measures aimed at reducing OC parental ties. For instance, the simulation will be able to model cases in which court orders limit the contacts between people involved in OC and their families or cases of OC members' conviction and imprisonment. In such cases, the ABM will temporarily decrease the relation that OC members have with their families and children while also providing juveniles at risk and their mothers with social and welfare support (e.g. school support, employment support).\\
The secondary socialisation policy aims at children and young people aged 6-18 who are in school. Crime-prone children, i.e. those with higher criminal propensity $C$, will be targeted with increased social support and/or increased welfare support. Increased social support may include: (a) better educational support (in the ABM, the agent will complete high school and/or achieve a higher level of education), (b) support of psychologists and social workers (promotion of pro-social relations and inhibition of anti-social relations, randomly creating friendship ties with non-deviant peers and adults), (c) increased social activities between children (children will randomly create new friendship ties), and/or (d) move child to new school classes. Increased welfare support instead may include providing a job to the child's mother (resulting in diversification of mother's networks) and/or providing the child with a job when they turn 16 (resulting in diversification of the child's social networks and lower risk of crime commission).

\subsection{Law Enforcement Targeting Policies}
This scenario aims at analysing the impact on the recruitment into OC of different law enforcement strategies in tackling OCGs. In particular, the possible targets will consist of (a) OC group bosses/lieutenants (in the model, potentially OC agents with high scores in measures such as betweenness) and (b) workers in “facilitator” positions. “Facilitators” include logistic workers, such as long-distance truck drivers and airport workers, and legal and financial advisors. These agents have increased opportunities for crime due to their work position (e.g. drug smuggling, money laundering). This scenario will put to the test and compare the effectiveness of LEAs targeting OC bosses/leaders and facilitators toward reducing crime rates and OC recruitment, both as separate and combined policies.\\
Regarding highly central OC members, two policies are proposed: (a.1) Higher scrutiny of OC members (it will decrease their ability to commit crimes and consequently create OC ties) and (a.2) Higher repression of OC members (it will lead to higher imprisonment rates for OC members). For what concerns facilitator workers, policies are related to: (b.1)	Higher scrutiny of facilitator positions (will decrease facilitators’ probability of crime commission and OC tie creation) and (b.2) Higher repression of criminal facilitators (will lead to higher imprisonment rates for criminal facilitators).

\section{Conclusions}

This paper has presented the rationale, input data and structure of an ABM designed to model recruitment into OC and to test different types of policies to contrast the strength of criminal groups in attracting new individuals. The model relies on empirical data gathered from official statistics when possible, and on information gathered from scientific literature otherwise. It is highly dependent on two dimensions, namely $C$ and $R$. $C$ formalises the individual probability of committing a crime of each agent, given its social, economic and criminal characteristics. $R$, conversely, quantifies the extent to which an agent is embedded into an highly OC-prone local community. This two-fold structure is a flexible and convenient solution that also permits to model and monitor non-criminal social processes, avoiding an overly narrow landscape of the simulation.  This ABM originates in the context of PROTON, a three-year Horizon 2020-funded project and will ultimately lead to the development of a user-friendly interactive tool at the disposal of policy-makers, practitioners and analysts interested in running virtual experiments to assess the potential consequences of a given policies. While this paper only presents the Palermo scenario, another model based on the same principles and structure will mirror the scenario of the Dutch city of Eindhoven, to assess external validity of results outside of the Italian context which is extremely peculiar due to the longstanding Mafia tradition. Furthermore, users will be able to test alternative societies via the possibility to modify sensitive hyper-parameters such as unemployment rate, criminal presence and law enforcement repression in order to resemble other European contexts. 
%
%
%
\bibliography{bib_oc.bib}

\begin{thebibliography}{10}
\providecommand{\url}[1]{\texttt{#1}}
\providecommand{\urlprefix}{URL }
\providecommand{\doi}[1]{https://doi.org/#1}

\bibitem{AbadinskyHistoryOrganizedCrime2014}
Abadinsky, H.: History of {Organized} {Crime} in the {United} {States}. In:
  Bruinsma, G., Weisburd, D. (eds.) Encyclopedia of {Criminology} and
  {Criminal} {Justice}, pp. 2204--2206. Springer New York (2014).
  \doi{10.1007/978-1-4614-5690-2_624},
  \url{http://link.springer.com/referenceworkentry/10.1007/978-1-4614-5690-2_624}

\bibitem{Arlacchimafiaimprenditrice1983}
Arlacchi, P.: La mafia imprenditrice. Il Mulino, Bologna (1983)

\bibitem{ArsovskaDecodingAlbanianorganized2015}
Arsovska, J.: Decoding {Albanian} organized crime: {Culture}, politics, and
  globalization. University of California Press, Oakland (2015)

\bibitem{AthanassaopolouOrganizedCrimeSoutheast2013}
Athanassaopolou, E.: Organized {Crime} in {Southeast} {Europe}. Routledge (Oct
  2013)

\bibitem{Brancaccioclandicamorra2017}
Brancaccio, L.: I clan di camorra: genesi e storia. Donzelli Editore (2017)

\bibitem{Brantinghamcomputationalmodelsimulating2005}
Brantingham, P.L., Glasser, U., Kinney, B., Singh, K., Vajihollahi, M.: A
  computational model for simulating spatial aspects of crime in urban
  environments. In: 2005 {IEEE} {International} {Conference} on {Systems},
  {Man} and {Cybernetics}. vol.~4, pp. 3667--3674 Vol. 4 (Oct 2005).
  \doi{10.1109/ICSMC.2005.1571717}

\bibitem{BruinsmaDifferentialAssociationTheory2014}
Bruinsma, G.: Differential {Association} {Theory} (2014)

\bibitem{BurgessDifferentialAssociationReinforcementTheory1966}
Burgess, R.L., Akers, R.L.: A {Differential} {Association}-{Reinforcement}
  {Theory} of {Criminal} {Behavior}. Social Problems  \textbf{14}(2),  128--147
  (Oct 1966). \doi{10.2307/798612},
  \url{https://academic.oup.com/socpro/article-lookup/doi/10.2307/798612}

\bibitem{CalderoniOrganizedCrimeLegislation2010}
Calderoni, F.: Organized {Crime} {Legislation} in the {European} {Union}.
  Springer International Publishing (2010),
  \url{http://www.springer.com/us/book/9783642043307}

\bibitem{CarringtonCooffendingCanadaEngland2013}
Carrington, P.J., van Mastrigt, S.B.: Co-offending in {Canada}, {England} and
  the {United} {States}: a cross-national comparison. Global Crime
  \textbf{14}(2-3),  123--140 (May 2013). \doi{10.1080/17440572.2013.787926},
  \url{http://www.tandfonline.com/doi/abs/10.1080/17440572.2013.787926}

\bibitem{DugatoMeasuringOrganisedCrime2019a}
Dugato, M., Calderoni, F., Campedelli, G.M.: Measuring {Organised} {Crime}
  {Presence} at the {Municipal} {Level}. Social Indicators Research  (Aug
  2019). \doi{10.1007/s11205-019-02151-7},
  \url{http://link.springer.com/10.1007/s11205-019-02151-7}

\bibitem{FijnautOrganisedCrimeEurope2004}
Fijnaut, C., Paoli, L.: Organised {Crime} in {Europe}: {Concepts}, {Patterns}
  and {Control} {Policies} in the {European} {Union} and {Beyond}. Springer
  Science \& Business Media, New York (Dec 2004)

\bibitem{Gottfredsongeneraltheorycrime1990}
Gottfredson, M.R., Hirschi, T.: A general theory of crime. Stanford University
  Press (1990)

\bibitem{GranovetterEconomicActionSocial1985}
Granovetter, M.: Economic {Action} and {Social} {Structure}: {The} {Problem} of
  {Embeddedness}. American Journal of Sociology  \textbf{91}(3),  481--510
  (1985), \url{http://www.jstor.org/stable/2780199}

\bibitem{KleemansCriminalCareersOrganized2008}
Kleemans, E.R., de~Poot, C.J.: Criminal {Careers} in {Organized} {Crime} and
  {Social} {Opportunity} {Structure}. European Journal of Criminology
  \textbf{5}(1),  69--98 (Jan 2008). \doi{10.1177/1477370807084225},
  \url{http://journals.sagepub.com/doi/10.1177/1477370807084225}

\bibitem{MallesonCrimereductionsimulation2010}
Malleson, N., Heppenstall, A., See, L.: Crime reduction through simulation:
  {An} agent-based model of burglary. Computers, Environment and Urban Systems
  \textbf{34}(3),  236--250 (May 2010).
  \doi{10.1016/j.compenvurbsys.2009.10.005},
  \url{http://www.sciencedirect.com/science/article/pii/S0198971509000787}

\bibitem{vanMastrigtCoOffendingAgeGender2009}
van Mastrigt, S.B., Farrington, D.P.: Co-{Offending}, {Age}, {Gender} and
  {Crime} {Type}: {Implications} for {Criminal} {Justice} {Policy}. British
  Journal of Criminology  \textbf{49}(4),  552--573 (Jul 2009).
  \doi{10.1093/bjc/azp021},
  \url{https://academic.oup.com/bjc/article-lookup/doi/10.1093/bjc/azp021}

\bibitem{McCarthyGettingStreetCrime1995}
McCarthy, B., Hagan, J.: Getting into {Street} {Crime}: {The} {Structure} and
  {Process} of {Criminal} {Embeddedness}. Social Science Research
  \textbf{24}(1),  63--95 (Mar 1995). \doi{10.1006/ssre.1995.1003},
  \url{https://ac.els-cdn.com/S0049089X85710034/1-s2.0-S0049089X85710034-main.pdf?_tid=25ea6aef-12fe-4d42-85a3-7e18ba82cd66&acdnat=1523278952_1c093a562932d2b2696093a8b3958106}

\bibitem{NardinSimulatingprotectionrackets2016}
Nardin, L.G., Andrighetto, G., Conte, R., Székely, A., Anzola, D.,
  Elsenbroich, C., Lotzmann, U., Neumann, M., Punzo, V., Troitzsch, K.G.:
  Simulating protection rackets: a case study of the {Sicilian} {Mafia}.
  Autonomous Agents and Multi-Agent Systems  \textbf{30}(6),  1117--1147 (Nov
  2016). \doi{10.1007/s10458-016-9330-z},
  \url{https://link.springer.com/article/10.1007/s10458-016-9330-z}

\bibitem{SavonaSystematicreviewsocial2017}
Savona, E., Calderoni, F., Superchi, E., Comunale, T., Campedelli, G.M.,
  Marchesi, M., Kamprad, A.: Systematic review of the social, psychological and
  economic factors relating to criminalisation and recruitment to {OC}. Tech.
  rep. (Dec 2017), \url{https://www.projectproton.eu/media-room/}

\bibitem{SkoganDimensionsDarkFigure1977a}
Skogan, W.G.: Dimensions of the {Dark} {Figure} of {Unreported} {Crime}. Crime
  \& Delinquency  \textbf{23}(1),  41--50 (1977).
  \doi{10.1177/001112877702300104},
  \url{https://doi.org/10.1177/001112877702300104}

\bibitem{StolzenbergCoOffendingAgeCrimeCurve2008}
Stolzenberg, L., D'Alessio, S.J.: Co-{Offending} and the {Age}-{Crime} {Curve}.
  Journal of Research in Crime and Delinquency  \textbf{45}(1),  65--86 (Feb
  2008). \doi{10.1177/0022427807309441},
  \url{http://journals.sagepub.com/doi/10.1177/0022427807309441}

\bibitem{SutherlandProfessionalThief1937}
Sutherland, E.H.: The {Professional} {Thief}. University of Chicago Press,
  Chicago (1937)

\bibitem{TroitzschExtortionRacketsEventOriented2016}
Troitzsch, K.G.: Extortion {Rackets}: {An} {Event}-{Oriented} {Model} of
  {Interventions}. In: Elsenbroich, C., Anzola, D., Gilbert, N. (eds.) Social
  dimensions of organised crime: modelling the dynamics of extortion rackets,
  pp. 117--131. Springer Science+Business Media, LLC, New York, NY, 1st edition
  edn. (2016)

\bibitem{VareseMafiasMoveHow2011}
Varese, F.: Mafias on the {Move}: {How} {Organized} {Crime} {Conquers} {New}
  {Territories}. Princeton University Press (Jan 2011)

\bibitem{WeermanTheoriesCooffending2014}
Weerman, F.M.: Theories of {Co}-offending. In: Bruinsma, G., Weisburd, D.
  (eds.) Encyclopedia of {Criminology} and {Criminal} {Justice}, pp.
  5173--5184. Springer New York, New York, NY (2014).
  \doi{10.1007/978-1-4614-5690-2_110},
  \url{https://doi.org/10.1007/978-1-4614-5690-2_110}

\bibitem{WilliamsNigerianCriminalOrganizations2014}
Williams, P.: Nigerian {Criminal} {Organizations}. In: The {Oxford} {Hanbook}
  of {Organized} {Crime}, pp. 254--269. Oxford University Press, New York
  (2014)

\end{thebibliography}
%

\end{document}